\documentclass[useAMS,usenatbib]{mnras}
\usepackage{pifont}
\usepackage{graphicx}
\usepackage{amssymb}
\usepackage{amsmath}
\usepackage{abbrevs}
\usepackage{xspace}
\usepackage[usenames,dvipsnames]{xcolor}
\begingroup
  \catcode`\_=\active
  \gdef_#1{\ensuremath{\sb{\mathrm{#1}}}}
\endgroup
\mathcode`\_=\string"8000
\catcode`\_=12

\newcommand{\Msolar}{M$_{\odot}$\xspace}

\newcommand{\atcc}{atoms/cm$^{3}$\xspace}

\newcommand{\simname}{\textsc}
\newcommand{\Stromgren}{Str\"{o}mgren\xspace}

\newcommand{\tick}{$\checkmark$}
\newcommand{\cross}{\ding{55}}

\newabbrev\ISM{Interstellar Medium (ISM)}[ISM]
\newabbrev\CSM{Circumstellar Medium (CSM)}[CSM]
\newabbrev\WNM{Warm Neutral Medium (WNM)}[WNM]
\newabbrev\WIM{Warm Ionised Medium (WIM)}[WIM]
\newabbrev\CNM{Cold Neutral Medium (CNM)}[CNM]
\newabbrev\IMF{Initial Mass Function (IMF)}[IMF]
\newabbrev\AMR{Adaptive Mesh Refinement (AMR)}[AMR]
\newabbrev\HGB{Horizontal Giant Branch (HGB)}[HGB]

\makeatletter
\renewcommand\maybe@space@{%
  \maybe@ictrue 
  \expandafter   \@tfor
    \expandafter \reserved@a
    \expandafter :%
    \expandafter =%
                 \nospacelist
                 \do \t@st@ic
  \ifmaybe@ic 
    \space
  \fi
}
\makeatother

\begin{document}

\title[Feedback in Clouds II: HII \& SN]{Feedback in Clouds II: UV Photoionisation and the first supernova in a massive cloud}
\author[S. Geen]
      {Sam Geen$^{1}$, Patrick Hennebelle$^{1}$, Pascal Tremblin$^{2}$, Joakim Rosdahl$^{3}$\\
{$^{1}$ Laboratoire AIM, Paris-Saclay, CEA/IRFU/SAp - CNRS - Universit\'e Paris Diderot, 91191, Gif-sur-Yvette, France}\\
{$^{2}$ Maison de la Simulation, CEA, CNRS, Univ. Paris-Sud, UVSQ, Universit\'e Paris-Saclay, 91191 Gif-sur-Yvette, France}\\
{$^{3}$ Leiden Observatory, Leiden University, P.O. Box 9513, 2300 RA, Leiden, The Netherlands}}
\date{\today}
\maketitle

\begin{abstract}

Molecular cloud structure is regulated by stellar feedback in various forms. Two of the most important feedback processes are UV photoionisation and supernovae from massive stars. However, the precise response of the cloud to these processes, and the interaction between them, remains an open question. In particular, we wish to know under which conditions the cloud can be dispersed by feedback, which in turn can give us hints as to how feedback regulates the star formation inside the cloud. We perform a suite of radiative magnetohydrodynamic simulations of a $10^5$ solar mass cloud with embedded sources of ionising radiation and supernovae, including multiple supernovae and a hypernova model. A UV source corresponding to 10\% of the mass of the cloud is required to disperse the cloud, suggesting that the star formation efficiency should be on the order of 10\%. A single supernova is unable to significantly affect the evolution of the cloud. However, energetic hypernovae and multiple supernovae are able to add significant quantities of momentum to the cloud, approximately $10^{43}$ g cm/s of momentum per $10^{51}$ ergs of supernova energy. This is on the lower range of estimates in other works, since dense gas clumps that remain embedded inside the HII region cause rapid cooling in the supernova blast. We argue that supernovae alone are unable to regulate star formation in molecular clouds, and that strong pre-supernova feedback is required to allow supernova blastwaves to propagate efficiently into the interstellar medium.
 
 \end{abstract}


\section{Introduction}
\label{introduction}

Massive stars release large quantities of energy into their environment. They produce protostellar jets, winds, radiation across a wide spectrum and supernovae. The first phase of stellar feedback occurs in dense molecular cloud environments in which the stars are born. In order for the energy from stars to propagate into the wider \ISM, it must first escape this cloud environment, either by destroying the cloud or creating sufficient channels through which the propagating shocks can escape.

In the previous paper, \cite{Geen2015b}, we determined a limit at which ionising radiation can escape molecular clouds using both numerical simulations and an analytic model. This model is based on arguments made in \cite{Tremblin2014a,Didelon2015}, which compare models of HII regions expanding into turbulent environments to observed HII regions. These models were constructed using previous analytic theory by \cite{KahnF.D.1954,SpitzerLyman1978,Whitworth1979,Franco1990,Williams1997,Hosokawa2006,Raga2012}.

In \cite{Geen2015b} we proposed a limit at which ionising photons are able to destroy their host cloud. This extends the argument of \cite{Dale2012}, who consider the case where the ionisation front cannot expand beyond the initial \Stromgren radius. We argue that if a calculated ``stall'' radius \citep{Draine1991} is smaller than the radius of the cloud, the ionisation front cannot escape the cloud. This in turn sets the ability for ionising radiation to regulate the environments in which stars form, and determines whether ionising radiation can suppress the star formation rate of molecular clouds.

Massive stars typically end their lives as supernovae \citep[for estimates of which stars become supernovae, see, e.g. ][]{Heger2003}. The evolution of the supernova remnant depends on the environment into which it expands. Understanding the momentum deposition from supernovae in star-forming environments is crucial to understanding processes in galaxies as a whole. Sub-grid models by, e.g., \cite{Hopkins2014,Kimm2015a} attempt to correct for a lack of numerical resolution by depositing a pre-calculated quantity of momentum around the supernova if the resolution is insufficient to resolve the blastwave properly \citep[see also ][ for a study of numerical limits on resolving supernova blastwaves]{Kim2015}. Analytic and 1D simulation work by \cite{Chevalier1974,Cioffi1988,Draine1991,Thornton1998,Haid2016} provides insights into this process, with simulations of supernova blastwaves by \cite{Iffrig2015,Kim2015,Martizzi2015,Kortgen2016} extending this to more complex environments using 3D numerical simulations. Supernovae shock against the surrounding medium, expanding adiabatically \citep{Sedov1946}. Eventually they reach a point where they begin to lose a significant fraction of their energy to radiative cooling \citep[see estimates by ][]{Cox1972}. After a longer period of time, the supernova remnant begins to merge with the surrounding medium \citep{Cioffi1988}.

Pre-supernova feedback as either stellar winds or ionising radiation has been found in simulations to enhance the final energy and momentum of the supernova remnant by injecting additional momentum and reducing the density of the environment into which the supernova occurs \citep{Dwarkadas2007,Fierlinger2015,Geen2015a}. \cite{Rogers2013,Walch2015} have had some success in driving outflows in simulations of molecular clouds with both supernova and pre-supernova stellar feedback. However, \cite{Draine1991} suggests that if the medium is sufficiently turbulent, the HII region will re-collapse before the supernova occurs, depending on the mass of the progenitor and the density of the surrounding medium. \cite{Krause2016} find that stellar feedback in very massive extragalactic clouds is ineffective at reducing the star formation efficiency. 


In this paper we explore the competition between pre-supernova ionising feedback and turbulence in molecular clouds, and the resulting evolution of the supernova remnant as it expands into the environment resulting from this competition. We simulate ionising radiation and supernovae in a turbulent cloud using \textsc{RAMSES-RT} \citep{Teyssier:2002p533,Fromang2006,Rosdahl2013}. The cloud is $10^5$ \Msolar, ten times more massive than the one studied in the previous paper. We choose this cloud mass because the slope of the cloud mass function ($\mathrm{d}N/\mathrm{d}M_c = M_c^{-1.7}$) means that more mass is expected to be found in clouds above $10^5$ \Msolar \citep[see review by ][]{Hennebelle2012}. Therefore, most of the stars in our Galaxy are expected to form in these clouds. It is thus important to study these objects if we wish to understand how feedback from massive stars interacts with both the host cloud and the wider Galactic \ISM.

We begin in Section \ref{methods} by presenting the simulations performed. We then extend the analysis of HII regions in the previous paper to a more massive cloud in Section \ref{beforesn}, and produce simple models that describe the time-dependence of the evolution of the ionisation front.  In Section \ref{aftersn}, we analyse the results of simulations that introduce a supernova into the cloud and HII region after the source of UV photons is extinguished, and how this compares to previous analytic and numerical theory. We extend the single supernova scenario to more energetic events in Section \ref{moresne}.


\section{Methods}
\label{methods}

\begin{table}
\centerline{\begin{tabular}{l c c c c c c c c c}
   \textbf{Name} & \textbf{log$_{10}$($S_{*}$/$\mathrm{s}^{-1}$)} & \textbf{Supernova?}\\
  \hline
 \simname{N00-NSN} & (no photons) & \cross & \\
 \simname{N49-NSN} & 49 &    \cross    &   \\
 \simname{N50-NSN} & 50 &    \cross    &   \\
 \simname{N51-NSN} & 51 &    \cross    &   \\
  \hline
 \simname{N00-SN}  & (no photons) & \tick &  \\
 \simname{N49-SN}  & 49 & \tick &  \\
 \simname{N50-SN}  & 50 & \tick &  \\
 \simname{N51-SN}  & 51 & \tick &  \\
 \simname{N50-HN} & 50 & Hypernova &  \\
 \simname{N50-MSN} & 50 & 10 $\times$ SN &  \\
  \hline
\end{tabular}}
  \caption{Table of the simulations included in this paper. ``\simname{N}'' refers to the number of photons deposited per second by the source, $S_*$, in all frequency groups in photons per second as a power of 10 (with ``\simname{N00}'' referring to a zero photon emission rate). ``NSN'' means that no supernova is included. ``SN'' means that a single $10^{51}$ erg supernova is included. ``HN'' means that a hypernova (modelled as a $10^{52}$ erg blast) is included. ``MSN'' means that ten supernovae are included, with $10^{51}$ ergs every 0.1 Myr. See Section \ref{methods} for more details about the simulation setup.}
\label{methods:simtable} 
\end{table}

We use the radiative magnetohydrodynamics code \textsc{RAMSES-RT} \citep{Teyssier:2002p533,Fromang2006,Rosdahl2013}. The system is described by an isolated turbulent, magnetised, self-gravitating initially spherical cloud placed at the centre of the simulation volume. After 2.53 Myr (one free-fall time for the cloud as a whole) we turn on a constant source of ionising UV photons in the centre of the simulation volume. After 3 Myr we turn off the source of photons and inject a thermal blast representing a supernova. More details on each component are given in the following sub-sections. Table \ref{methods:simtable} lists all the simulations used in this paper.

\subsection{Initial Conditions}
\label{methods:ics}

In this simulation we consider one set of initial conditions only. For a theoretical description of the effect of varying cloud properties on the shape of HII regions, see \cite{Geen2015b}. These initial conditions are similar to the setup given in \cite{Iffrig2015,Geen2015b,Lee2016a}.

The simulation volume is a cubic box of length 86.3 pc. This volume is divided into 256 cells on a side. We allow a further 2 levels of refinement (i.e. subdivision of a single cell into 8 cells), giving 1024 cells on a side effective resolution. The simulation thus has a minimum spatial resolution of 0.33 pc everywhere and 0.084 pc in the most refined regions. At all times if a cell is found to be Jeans unstable it is allowed to refine up to the maximum level. Additionally, shortly before the supernova is launched we fully refine the central 1.5 pc of the simulation volume in order to capture the shock evolution properly. For \simname{N51-SN} and \simname{N51-NSN} we run identical simulations with twice the box length but identical spatial physical resolution and refinement criteria in order to follow shocks that would otherwise escape the box.

\begin{figure*}
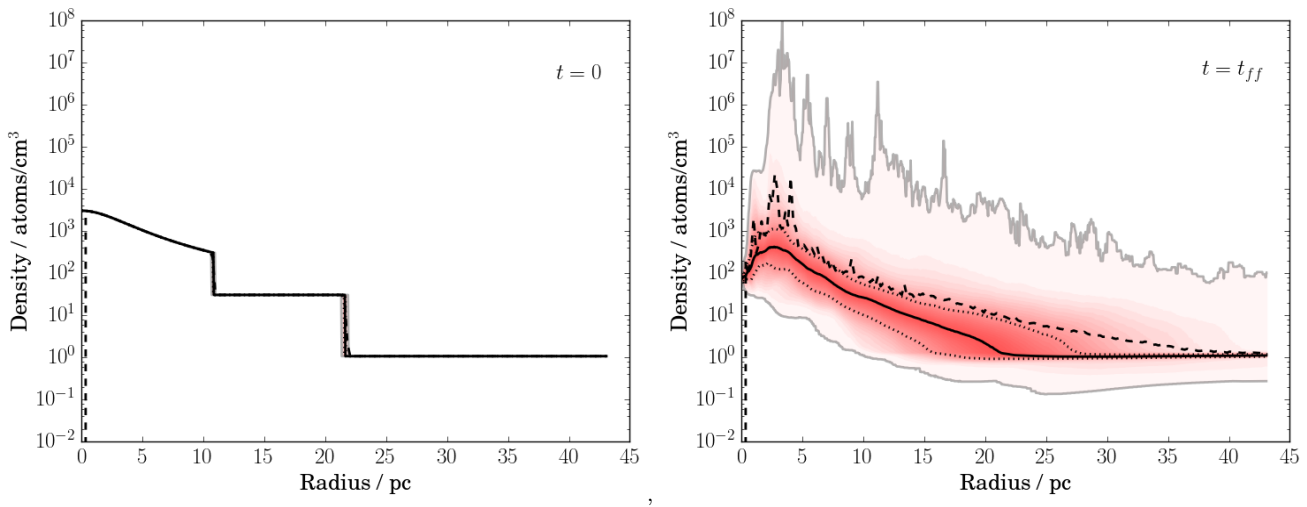

\centerline{\includegraphics[width=0.48\hsize]{plots/fig1a.pdf},
\includegraphics[width=0.48\hsize]{plots/fig1b.pdf}}
\caption{Density profile of the cloud. The plot on the left gives the profile at the start of the simulation, while the plot on the right is taken after one free-fall time, just before the HII source is turned on. We cast a set of rays from the source position to the edge of the simulation volume in evenly spaced directions as described in Appendix E of \protect\cite{Geen2015b}. Along each ray we sample the density profile. The minimum and maximum values at each radius are given as a thick grey line. The median value at each radius is given as a solid black line, while the mean is a black dashed line. The interquartile range is bounded by black dotted lines. Each second percentile of probability from the median value is shaded from red to white, with the closest percentiles to the median shaded as red and the furthest as white.}
 \label{beforesn:densprofiles}
\end{figure*}

The mass of the cloud studied in this paper is set to $10^{5}~$\Msolar. We impose a spherically symmetric density profile onto the simulation volume (see the left panel of Figure \ref{beforesn:densprofiles}). This is given by $n(r,t=0) = n_0 / (1 + (r/r_c)^2)$ for hydrogen number density $n$ at radius $r$ and time $t=0$ with peak density $n_0=2850$ \atcc and characteristic radius $r_c = 3.6$ pc. We impose a cut-off at $3~r_c$ (where $n(3 r_c,t=0) = 0.1~n_0$) Outside this radius, a uniform density field of density 20 \atcc is imposed out to 21 pc. Beyond this, the density field is a uniform 1 \atcc. The cloud has a global free-fall time $t_{ff}=2.53$ Myr, defined as $3 \sqrt{\frac{3 \pi}{32 G n_0 m_H/X}}$, where $m_H$ is the mass of a hydrogen atom and $X=0.76$ is the hydrogen mass fraction. The temperature inside the inner part of the cloud is set to 10 K, while the temperature in the medium outside the cloud is set to 4000 K. The magnetic field is initially 25 $\mu$G in the centre of the cloud and 4.2 $\mu$G outside, chosen such that the ratio between the free-fall time and the Alfv\'en crossing time is 0.2. Note that the magnetic field strength increases as the cloud evolves.

A turbulent velocity field is imposed over the grid, such that the kinetic energy in turbulence in the cloud is approximately equal to the gravitational energy of the cloud. The turbulence has a Kolmogorov power spectrum (i.e. $P(k) \propto k^{-5/3}$) with random phases.

\subsection{Radiative Transfer}
\label{methods:rt}

 \textsc{RAMSES-RT} \citep{Rosdahl2013} uses a first-order moment method for the advection of photons, closing the set of equations with the local M1 expression for the radiation pressure tensor. It tracks the ionisation states of hydrogen and helium in the gas, and couples the interactions between the photons and the gas on-the-fly. We split the radiation into three groups, bracketed by the ionisation energies of HI, HeI and HeII (13.6, 24.6 and 54.2 eV for the lower bounds of each). In all of the simulations in this paper we assume a Solar metallicity everywhere at all times. We do not include photons below the ionisation energy of hydrogen, nor do we include radiation pressure \citep[as in ][]{Rosdahl2015}. A reduced speed of light of $10^{-4}$ c (= 30 km/s, or 2.4 $c_i$) is used. This is in order to prevent the timestep becoming prohibitively short. We chose the minimum value such that the speed of ionisation fronts in our simulations is the same as that for a larger speed of light.

\subsection{Radiative Cooling}
\label{methods:physics}

In each cell, radiative cooling and heating of hydrogen and helium is performed by the radiative transfer module of \textsc{RAMSES-RT} as in \cite{Rosdahl2013}. For metals, we have two cooling and heating functions. The first, ``neutral'' function $N(T)$, is the prescription of \cite{Audit2005}, which includes carbon, oxygen and dust grains as well as the ambient UV background in the \ISM, transitioning to the prescription of \cite{Sutherland1993} above $10^4$ K. 

A further ``photoionised'' cooling function $P(T)$ is included, which is a piecewise fit to \cite{Ferland2003} \citep[see also ][]{Osterbrock1989}. This fit has a constant $3\times10^{-24}$ ergs cm$^3$/s below 9000 K and $2.2\times10^{-22}$ ergs cm$^3$/s above $10^5$ K, interpolating between these two points inside these temperatures. The metal cooling rate is set to $N(T)$ if $N(T) > P(T)$, where a positive value indicates cooling rather than heating. Otherwise, the cooling and heating rate is given by $x P(T) + (1-x) N(T)$, where $x$ is the hydrogen ionisation fraction.


\subsection{UV Source Properties}
\label{methods:uv}

We implement UV radiation in a similar way to \cite{Geen2015b}. The source of UV photons is modelled as a point source of ionising photons in the three photon groups. For each source we calculate a photon energy and emission rate assuming black body emission. Frequency-dependent cross sections are taken from \cite{Verner1996} via \cite{Hui1997}. In principle, the spectrum from an OB star will differ from a blackbody spectrum, but in practice we find that the exact spectrum of ionising photons is of secondary importance provided that the number of hydrogen-ionising photons is the same \citep[see also ][]{Haworth2015}.

We use three sources of ionising photons in this paper, as well as control simulations with no ionising radiation. These sources have ionising photon emission rates $S_*$ of $10^{49}~$s$^{-1}$, $10^{50}~$s$^{-1}$ and $10^{51}~$s$^{-1}$. In order to compare these values to physical sources, we use the results of \cite{Vacca1996} and Starburst 99 \citep{Leitherer2014}. The sources are taken to be, respectively, a 40 \Msolar star, a 100 \Msolar star, and a cluster of ten 100 \Msolar stars. Alternatively, the hydrogen-ionising photon emission rates correspond to clusters of masses 100, 1000 and 10000 \Msolar respectively, i.e. 0.1\%, 1\% and 10\% of the total mass of the cloud ($10^5$ \Msolar). See Appendix \ref{appendix:clusteremission} for a calculation of these values. Each of these sources are turned on for 3 Myr, at which point the stars enter the \HGB phase and stop producing significant quantities of ionising photons.


\subsection{Supernova Model}
\label{methods:sn}

After 3 Myr, when the UV photons are extinguished, we launch the supernova. This is a simplification of the full stellar lifecycle, which is beyond the scope of this paper. Instead, the intent of this paper is to use semi-realistic prescriptions that allow us to explore the behaviour of feedback in molecular clouds using controlled conditions. We discuss more sophisticated models in Section \ref{discussion}.

We implement the supernova as a point injection of thermal energy into the centre of the simulation volume, at the same location as the source of UV radiation. Our spatial resolution (0.084 pc around the supernova) satisfies the resolution criterion given by \cite{Kim2015} (0.14 pc at $10^4$ \atcc). In our primary supernova model (labelled ``SN''), we inject $10^{51}$ ergs of energy and 1 \Msolar of mass, representing the ejecta from the supernova. We also include a hypernova model \citep[labelled ``HN'', ][]{Nomoto2005}, which is identical to the ``SN'' model except that we inject $10^{52}$ ergs and 10 \Msolar. A further model, labelled ``MSN'', includes ten supernovae with energy $10^{51}$ ergs and 1 \Msolar of ejecta, launched 0.1 Myr apart \citep[see ][ for a similar model]{Kim2015}.

\section{The HII Region Before the Supernova}
\label{beforesn}

In this section we present the results of our simulations containing HII regions before the supernova is launched. This includes both the phase before the source of ionising photons is turned on and the period during which the source is shining. We then review predictions of the behaviour of the ionisation front made in \cite{Geen2015b} and extend these models to describe the time-dependent behaviour of the system.

\subsection{The Prestellar Phase}
\label{beforesn:prestellar}

The cloud initially evolves magneto-hydrodynamically under self-gravity. The cloud fragments over roughly one free-fall time. We plot the density profile at the start of the simulation and after one free-fall time in Figure \ref{beforesn:density_profile}. The dense clumps formed after one free-fall time cover only 2\% or less of the solid angle around the source. For this reason, the median density profile is significantly lower than the mean density profile.

As the turbulence dissipates, the cloud begins to contract, with dense clumps falling towards the centre of the cloud. The dissipation time is on the order of the crossing time for the velocity dispersion of the cloud, a few times the free-fall time of the cloud \citep{Lee2016}. This effect was also seen in \cite{Geen2015b}. 

The densest clumps fall towards the centre of the cloud. They may be prevented from doing so by radiative feedback, though the precise behaviour of the clumps is complex and difficult to quantify using a simple model. Since the densest gas in the cloud is found in these clumps, if they are able to reach the position of the source they will quench the HII region and cause it to ``flicker'' \citep{Peters2010}.

\subsection{Expansion of the HII Region}
\label{beforesn:expansion}

Once the stars begin to radiate they ionise the gas around them. This photoionised gas has an equilibrium temperature around $10^4$ K, set by the radiative cooling function. The ionisation front reaches a hydrostatic limit called the \Stromgren radius, proportional to $S_*^{1/3}$, where all of the photons are used to keep the gas photoionised. Due to a pressure difference with the neutral gas, the ionisation front expands while maintaining photoionisation equilibrium. In a uniform medium with no source of external pressure, the ionisation front radius $r_i$ expands such that $\mathrm{d}r_i/\mathrm{d}t$ is proportional to $S_*^{1/7}$ \citep{Matzner2002}.

Eventually, the front ``stalls'' (i.e. is unable to expand further) due to ram pressure from external turbulence. In \cite{Geen2015b} we present a model that estimates the radius at which this occurs, $r_{stall}$, for a virialised cloud with radius $r_{cloud}$ (see Appendix \ref{appendix:expansion}). We find three regimes governed by the ratio $r_{stall} / r_{cloud}$. If this ratio is much larger than 1, the cloud is dispersed. If it is much smaller than 1, the ionisation front is trapped by the cloud. If the ratio is close to 1, the ionisation front escapes as a blister region but does not completely destroy the cloud.

The value of $r_{stall}$ depends on the density profile of the cloud. In \cite{Geen2015b} we fit a single spherically-averaged radial power law to the density field in the simulated cloud, in addition to sampling the density of the simulated cloud as a function of radius and time. The single power law fit provided an adequate match to the simulation results. However, in this paper, where we use a more massive cloud, we find that we must adopt the two-phase fit to the density profile of the cloud given by \cite{Franco1990}, with a flat cloud core surrounded by a power law density field. This is given by
\begin{equation}
\begin{split}
 n_{ext}(r <= r_c) &= n_c \\
 n_{ext}(r > r_c) &= n_c (r/r_c)^{-w}
 \end{split}
 \label{beforesn:density_profile}
\end{equation}
where $r_c=~$3.6 pc is the scaling radius given in the initial conditions and $n_c$ is the density at $r=r_c$. These values are found by fitting the spherically-averaged mean density field outside $r_c$ to a power law with free parameters in $n_c$ and $w$ at the time the source is turned on. The fit gives $n_c=~$1612 \atcc and $w=2.93$, i.e. with a very steep transition from the cloud core to the diffuse medium outside.

There is some flexibility in deciding the optimal fit for the full 3D density profile. The fit to Equation \ref{beforesn:density_profile} is degenerate depending on the value of $r_c$ chosen. In addition, in cases where a large quantity of mass is in small clumps sufficiently far from the position of the source, the median density profile offers a better fit to the effective density field experienced by the HII region.

Since the power law slope at $r > r_c$ is so steep, we argue that if $r_{stall}$ exceeds $r_c$, the ionisation front is able to escape the cloud. This is equivalent to our limit comparing $r_{stall}$ and $r_{cloud}$ in \cite{Geen2015b}.

\subsection{Comparison to the Simulations}
\label{beforesn:comparison}

We calculate the ratio $r_{stall}/r_{c}$ for each of our simulations. The values for \simname{N49-NSN}, \simname{N50-NSN} and \simname{N51-NSN} are 0.82, 1.1 and 1.6. Each simulation is thus in (or close to) a regime in which we predict the front should stall, almost escape the cloud or expand more or less freely. Again, there is some error in these estimates depending on the quality of the fit of a spherical density profile to the simulations.

\begin{figure*}
\centerline{\includegraphics[width=0.48\hsize]{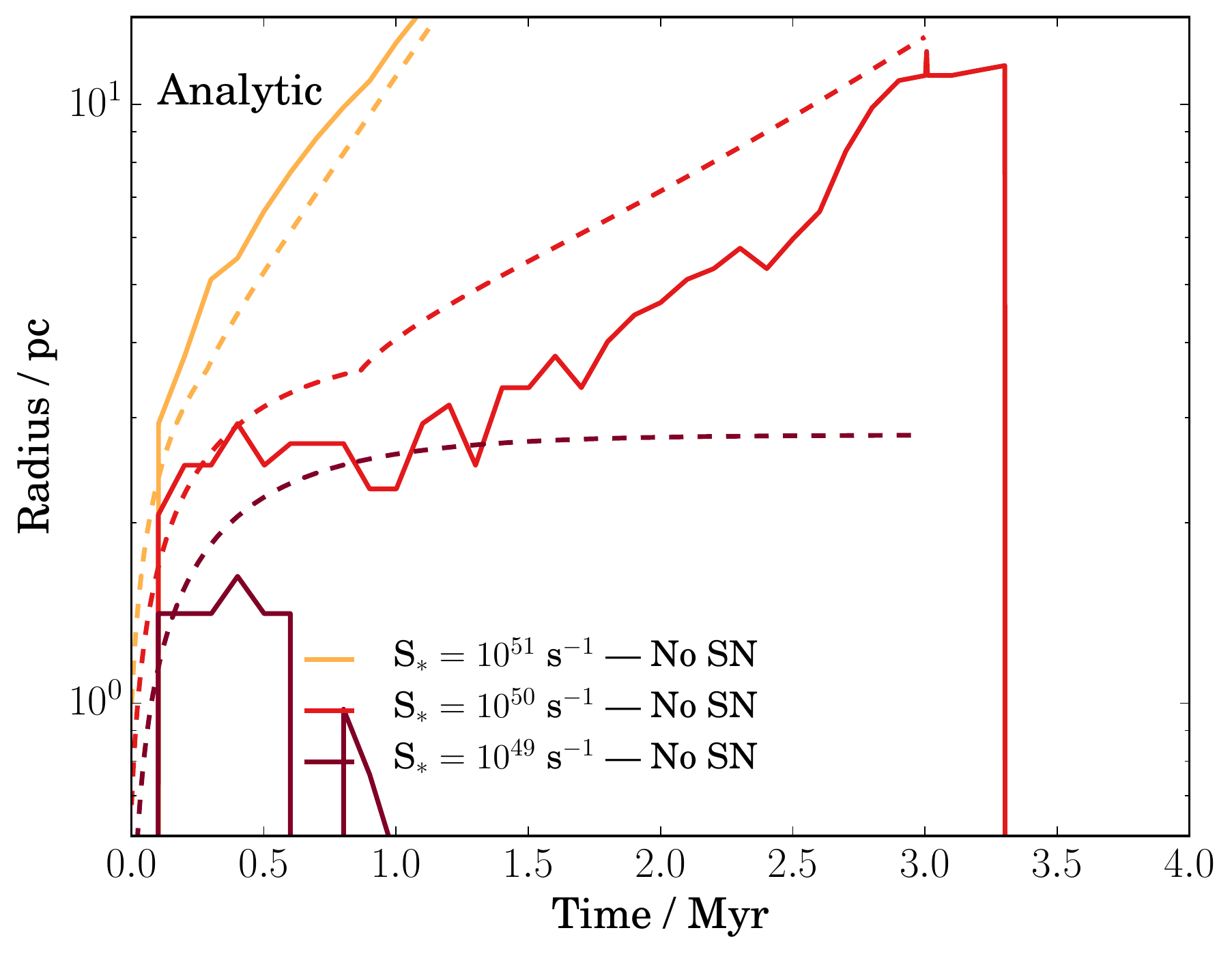}
\includegraphics[width=0.48\hsize]{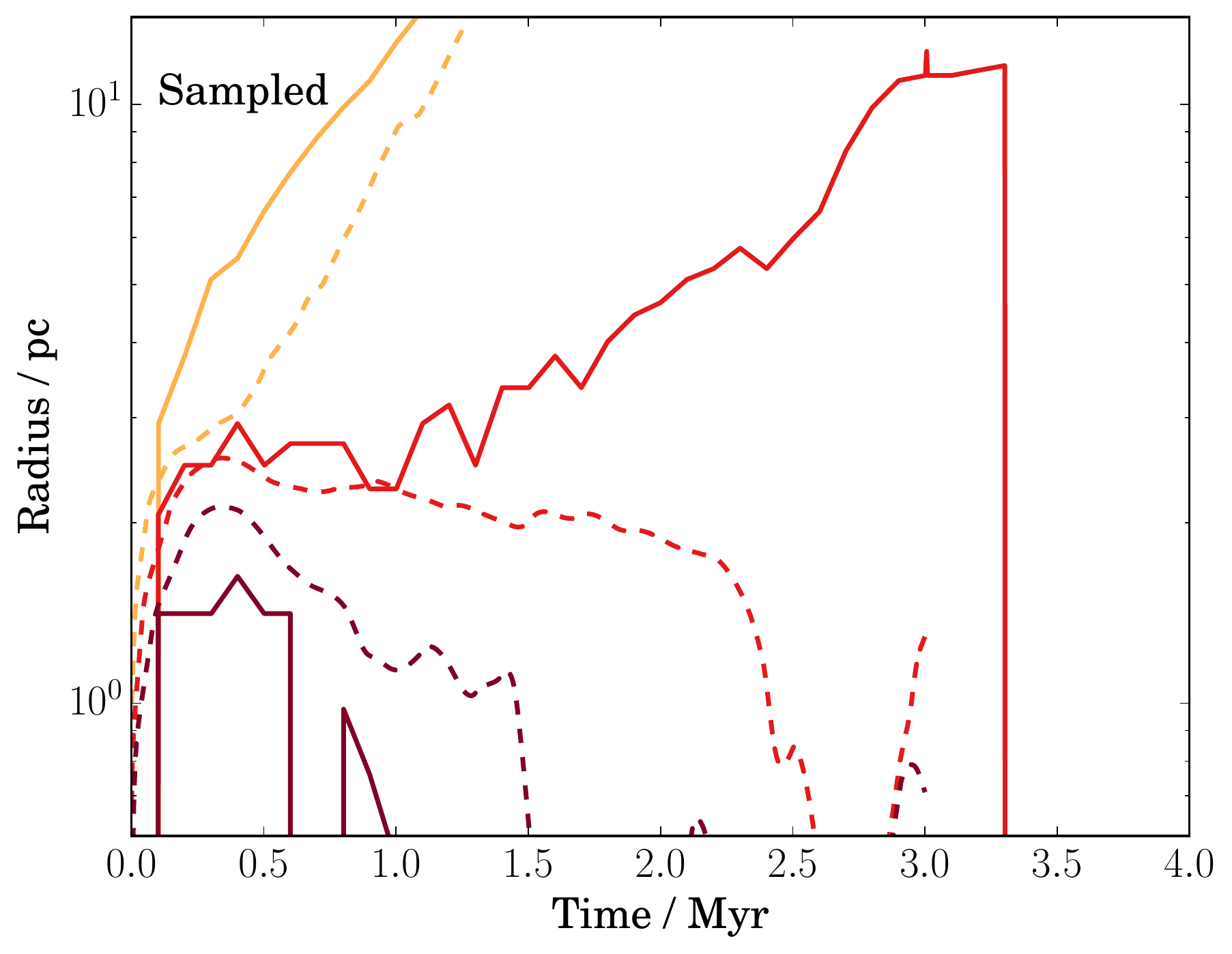}}
\caption{Comparison of our simluation results (solid lines) and model predictions (dashed lines) for the median radial expansion of the ionisation front with various photon emission rates. The left plot labelled ``Analytic'' uses the density profile fit described in Equation \ref{beforesn:density_profile}. The plot on the right labelled ``Sampled'' samples the density and radial velocity field in simuation \simname{N00-NSN} to predict the behaviour of the ionisation front. See Section \ref{beforesn:comparison} for more details.}
\label{beforesn:ragacompare}
\end{figure*}

In Figure \ref{beforesn:ragacompare} we repeat the calculation performed in Figure 8 of \cite{Geen2015b}, solving Equation \ref{expansion:raga_like} numerically using the density field given in Equation \ref{beforesn:density_profile} to estimate the expansion of the ionisation front. This model includes ram pressure from turbulent motions in the cloud, which we assume to be virialised. In \cite{Geen2015b} we set $v_{ext}$ to the escape velocity whereas here we use $v_{ext} = -v_{vir}(r) = -\sqrt{\frac{6G M(<r)}{5 r}}$, which is 77 \% of the escape velocity. Using the escape velocity instead results in a smaller value for $r_{stall}$. This model is labelled ``Analytic''.

We also  sample the spherically-averaged density and radial velocity field at each time and radial position in simulation \simname{N00-NSN} and solve Equation \ref{expansion:raga_like} using these inputs. We label this solution ``Sampled''. We plot both these models against the median radius of the ionisation front in our simulations at each timestep. Gas in the simulations is assumed to be ionised if its hydrogen ionisation fraction is above 0.1.

The behaviour of the ionisation fronts in each of the simulations agrees well with the ``Sampled'' models, with the exception of the $10^{50}~$s$^{-1}$ simulation, whose ionisation front does not collapse as the model predicts. This is due to the fact that the dense clumps in this simulation are prevented from reaching the centre of the cloud, and hence the photons are able to escape over the majority of the lines of sight around the source.

The expansion in the ``Analytic'' model follows the broad features of the ionisation fronts with $10^{50}$ and $10^{51}$ s$^{-1}$ sources as they expand. The expansion of the Analytic model is too fast in the $10^{49}$ and $10^{50}$ s$^{-1}$ sources, suggesting either that our choice of density is too low or that the velocity of the gas is higher than our estimate. In addition, since we assume a static density field we do not capture the collapse of the ionisation front in the $10^{49}$ s$^{-1}$ case. We present a term correcting for this collapse in Appendix \ref{appendix:collapsing}.

The equations governing the expansion of the ionisation front in a turbulent medium are non-linear. For this reason in \cite{Geen2015b} we solve these equations numerically. An analytic limit at which the front stalls is also given. It is useful to know over what time this stall radius is reached. In Appendices \ref{appendix:timevolution} and \ref{appendix:collapsing} we show that this timescale over which the front either stalls (or collapses, if the cloud is strongly accreting) is roughly equal to the free-fall time in the cloud core. This suggests that star formation in the cloud and the cloud destruction occur over similar timescales, making it difficult to find an accurate analytic model for the point at which star formation is frozen out by the destruction of the cloud via radiative feedback.

In general, the models obey the broad behaviour defined by our limit in $r_{stall}$, with some uncertainty due to the complex structure of the cloud as the ratio $r_{stall}/r_c$ approaches 1 and the stalled front transitions to a freely expanding front. For the cloud simulated here, this means that the ionising photons are only able to significantly disperse the cloud with a $10^{51}$ s$^{-1}$ source, though the $10^{50}$ s$^{-1}$ source is able to drive a large ionised bubble that escapes the cloud in certain directions but does not escape the simulation volume.



\subsection{After the Source is Extinguished}
\label{beforesn:additional}

After a few million years the star no longer produces a significant amount of ionising photons. This causes the HII region to cool on a timescale governed by the density of the gas inside the HII region. We discuss a simple model for this in \cite{Geen2015a}, where the ionisation front is no longer in ionisation equilibrium. Once the source of pressure from photoheated gas is reduced, the shell around the ionisation front can continue to expand due to momentum conservation, as derived by \cite{Hosokawa2006}.

In general, both of these cases are most important when the ionisation front is able to escape the cloud. If the front is stalled or collapses, the residual momentum will be close to zero. Indeed, once the source is turned off we observe a gradual re-collapse of the cloud.

Our results agree with the conclusions of \cite{Geen2015b}. If the source is strong enough to resist the infall of clumps, the ionisation front can expand into the external medium. Otherwise, it stalls or contracts. The boundary between these two cases is found at around $r_{stall} = r_{c}$.

\section{After the Supernova}
\label{aftersn}

So far we have discussed the evolution of the cloud and HII region up to the point where the source is extinguished, equivalent to the point at which the most massive star in the cluster reaches the end of its life. In this section we discuss the phase of our simulations after this star goes supernova.

\subsection{The Structure of the Cloud Before and After the Supernova}
\label{aftersn:justbefore}

We first review the properties of the cloud at the point where the supernova is injected for the different photon emission rates where $r_{stall}/r_{c}$ is less than, roughly equal to and greater than 1.

\begin{figure*}
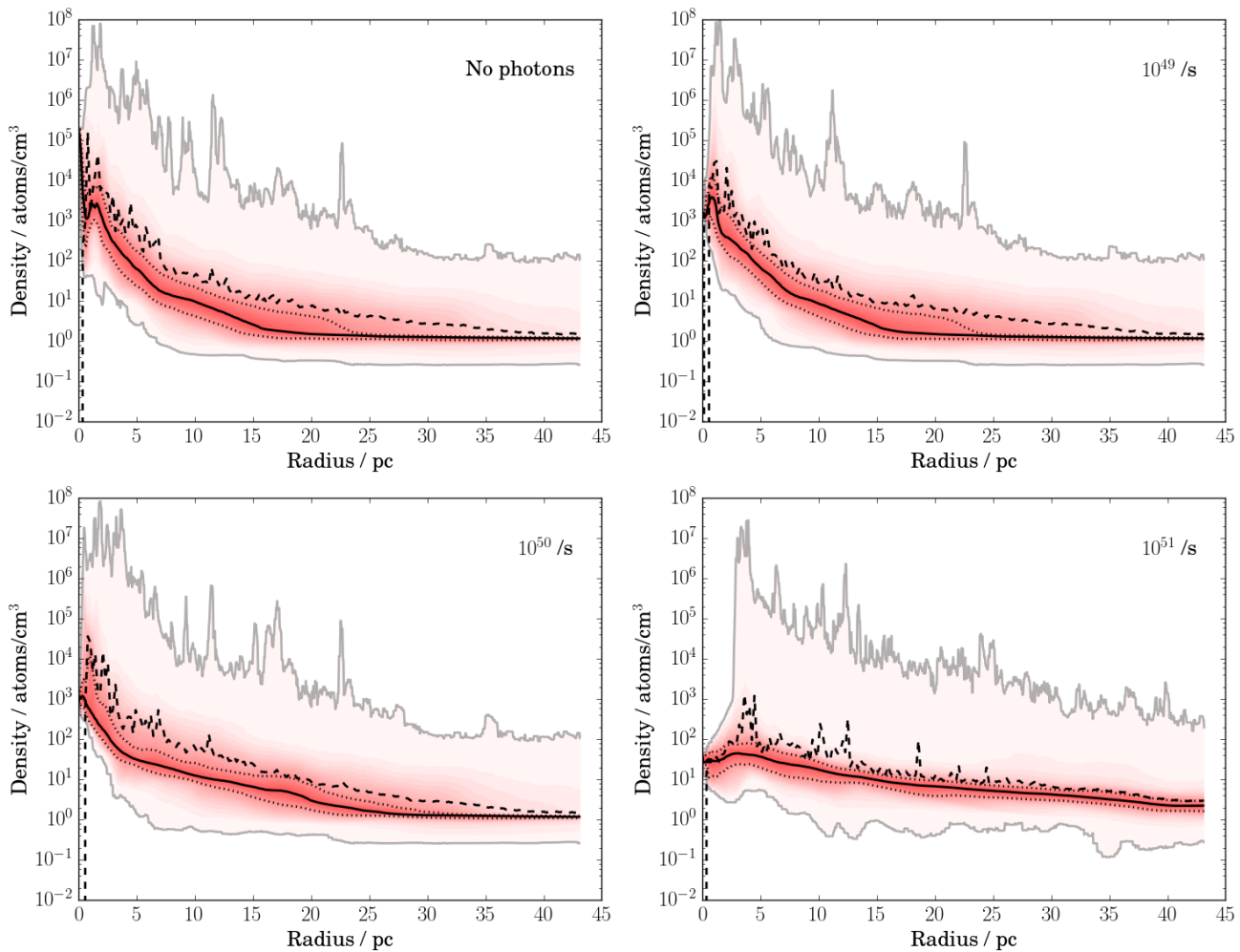

\centerline{\includegraphics[width=0.48\hsize]{plots/fig3a.pdf}
\includegraphics[width=0.48\hsize]{plots/fig3b.pdf}}
\centerline{\includegraphics[width=0.48\hsize]{plots/fig3c.pdf}
\includegraphics[width=0.48\hsize]{plots/fig3d.pdf}}
\caption{Density profiles just before the supernova is launched for each of the photon emission rates. Top left: no source. Top right: $10^{49}$ photons/s. Bottom left: $10^{50}$ photons/s. Bottom right: $10^{51}$ photons/s. The plots are constructed as in Figure \ref{beforesn:densprofiles}.}
 \label{aftersn:densprofiles}
\end{figure*}

In Figure \ref{aftersn:densprofiles}, we plot the density profiles around the source just prior to the supernova. Due to turbulent dissipation in the cloud and the infall of dense clumps, the profile becomes more peaked over time, particularly with the  $10^{49}$ photons/s source. Only the $10^{51}$/s source is able to significantly flatten the cloud profile. Similarly, as described in Section \ref{beforesn:comparison}, only the $10^{51}$/s source is able to drive ionised flows out of the simulation volume, while the ionised gas around the weaker sources remains bounded by neutral gas. This means that the supernova shock must break through this material if it is to move out into the external medium.

\begin{figure*}
\centerline{\includegraphics[width=0.98\hsize]{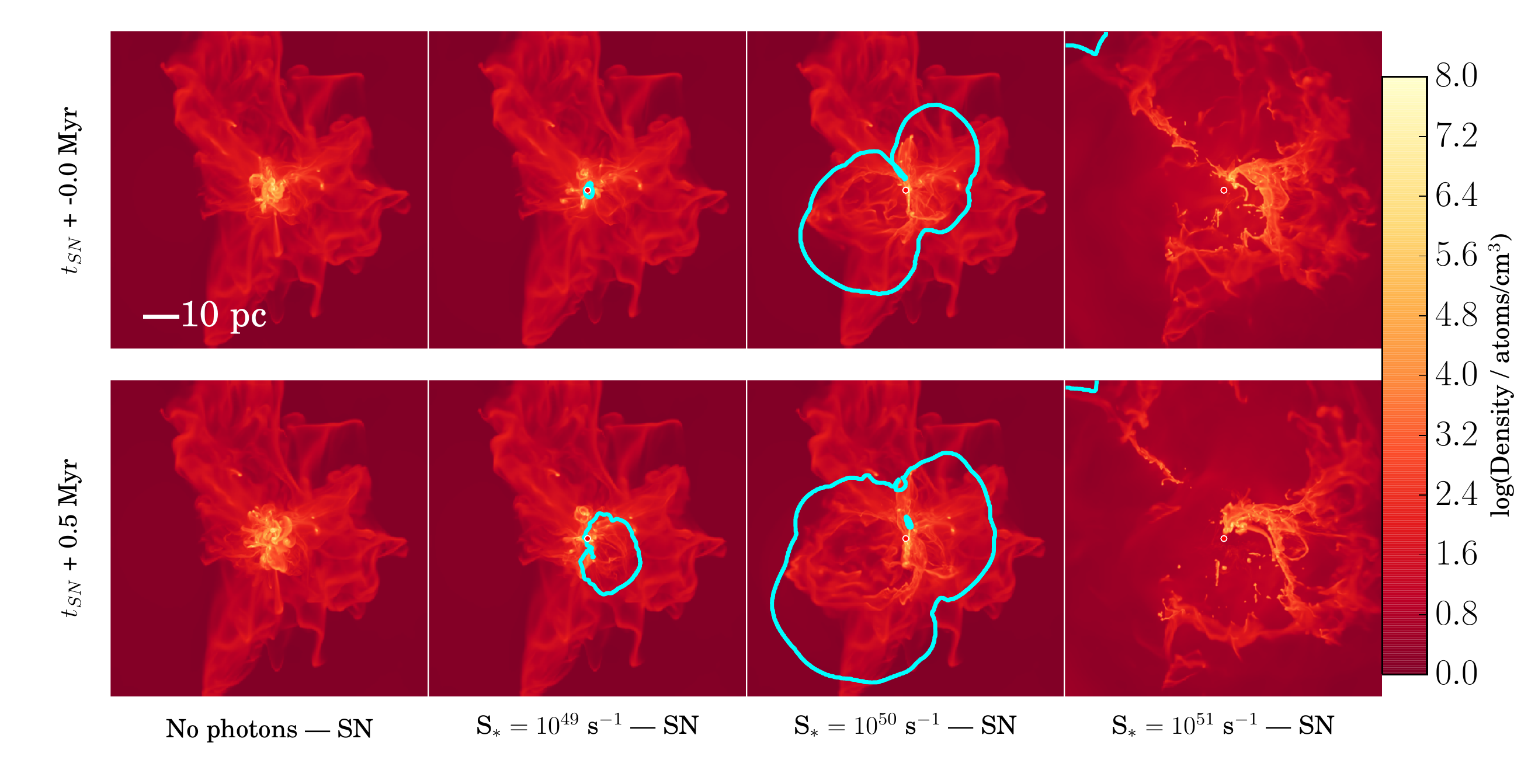}}
\caption{Images showing the maximum density along the line of sight for each of the simulations containing supernovae at the time the supernova is injected (5.52 Myr after the start of the simulation, and 3 Myr after the source of ionising radiation is turned on) and 1 Myr afterwards. The cyan line shows the maximum extent of ionised gas along the line of sight. This gas can be either photoionised in the case of the HII region or collisionally ionised in the case of the hot bubble inside the supernova remnant. The red dot shows the position of the source of ionising radiation where one is included (note that this source is turned off at the times shown in these images). The total length of image is 86.3 pc in all dimensions.}
\label{aftersn:pictures}
\end{figure*}
\begin{figure*}
\centerline{\includegraphics[width=0.98\hsize]{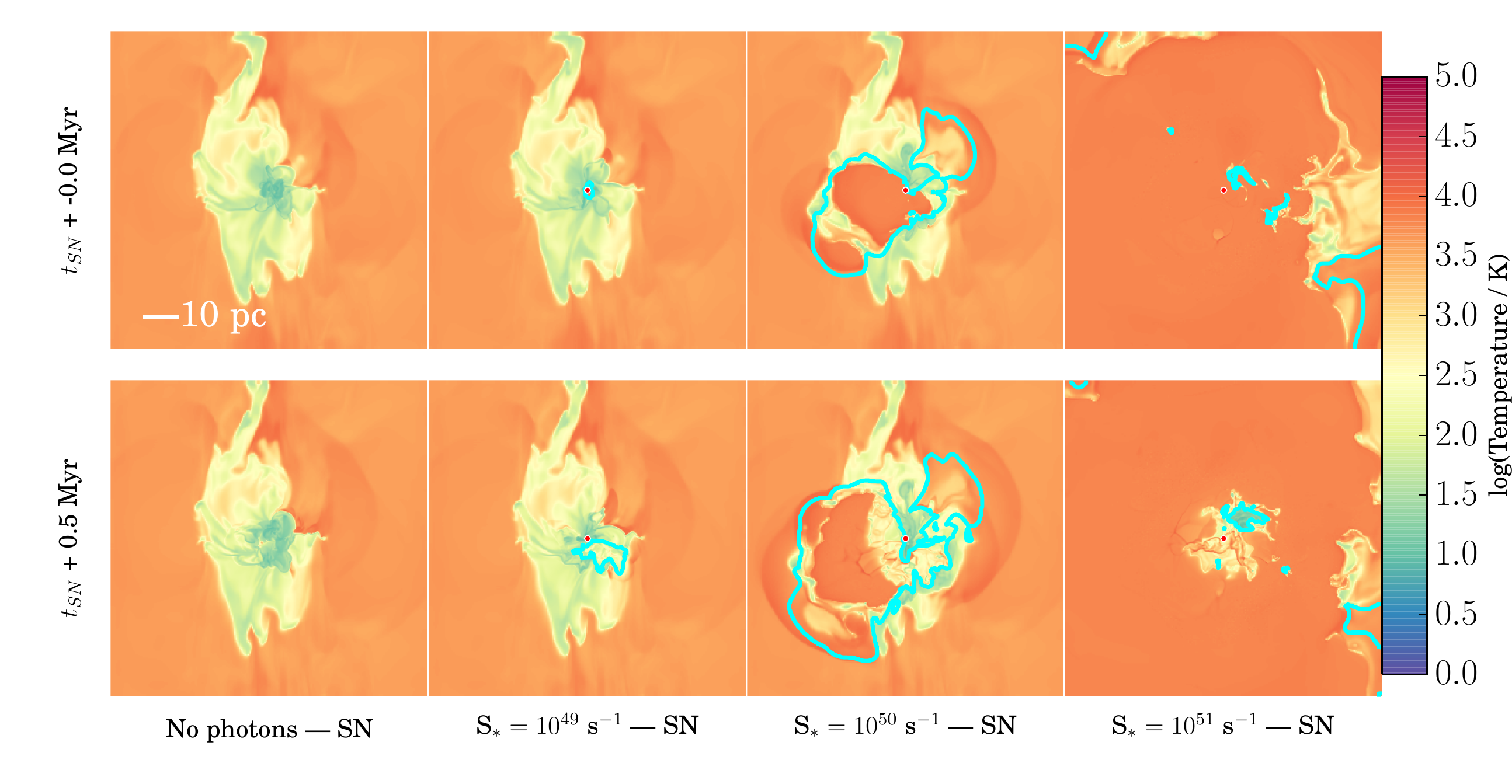}}
\caption{As in Figure \ref{aftersn:pictures} but showing gas temperature as a slice through the centre of the simulation volume.}
\label{aftersn:temperature}
\end{figure*}

Figure \ref{aftersn:pictures} contains projections of the density field before and after the supernova. As with the ionisation front, the supernova remnant expands preferentially through lower-density channels in the cloud. Since the $10^{51}$/s source was the only source capable of expelling a significant quantity of the cloud material -- 80\% as a function of solid angle around the source -- this is the only simulation in which the supernova successfully escapes the simulation volume and into the surrounding medium. In Figure \ref{aftersn:temperature} we plot a slice through the temperature field around the source. Since the supernova explodes in dense gas, the remnant has cooled to around $10^4$ K after a few hundred kyr. Even in the $10^{51}$ /s simulation, the temperature is reduced by reflection shocks of denser gas passing through the hot, diffuse bubble due to the asymmetric shape of the remnant and the recollapse of the cloud under gravity.

\subsection{Analytic Overview}
\label{aftersn:analytic}

The classical picture of the expansion of supernova remnants in a uniform medium is described by \cite{Chevalier1977}. Once the supernova blastwave leaves the surface of the star, there is an initially kinetic phase as the ejecta moves outwards ballistically. Once the mass displaced by the supernova is roughly equal to the supernova ejecta mass, the supernova ejecta shocks against the surrounding medium and enters the adiabatic, ``Sedov'' phase \citep{Sedov1946}. As the remnant expands, it sweeps up unshocked matter, forming a shell around the shock-heated gas. It also begins to cool radiatively, entering the pressure-driven snowplough phase. As radiative cooling becomes significant, the thermal pressure inside the shell drops to the point where the shell expands only due to momentum conservation in the momentum-conserving snowplough phase.

An equation for the time at which radiative cooling is complete and shell radius at which this occurs is given in Section c) ii) of \cite{Cox1972} as:
\begin{equation}
\begin{split}
 t_\Lambda &= 5.0~(\epsilon_0)^{4/17} n_0^{-9/17} \times10^4 \mathrm{years,} \\
 r_\Lambda &= 22.1~(\epsilon_0)^{5/17} n_0^{-7/17} \mathrm{pc} 
 \end{split}
 \label{aftersn:cox1972}
\end{equation} where $\epsilon_0$ is a unitless quantity given by the energy of the supernova divided by $10^{51}$ ergs and $n_0$ is a unitless quantity given by the density in the surrounding medium divided by 1 \atcc (assumed to be uniform). We find that for values typical in our simulation the supernova should exit the Sedov phase before it leaves the core of the cloud, with $r_\Lambda = 3.3~$pc at $n_0 = 100$ and $r_\Lambda = 0.50~$pc at $n_0 = 10000$.

\cite{Iffrig2015} find that an accurate prediction of the final momentum of the supernova remnant $p_\Lambda$ can be achieved by calculating the momentum of the remnant in the Sedov phase at $t_\Lambda$, where
\begin{equation}
p_\Lambda \propto n_0^{-2/17} {\epsilon_0}^{16/17}.
\label{aftersn:pc}
\end{equation} 
In other words, the final momentum is weakly linked to density (around 1 to 3 $\times 10^{43}~$ g cm/s in our cloud), and roughly proportional to the initial supernova energy. \cite{Iffrig2015} find that this simple argument compares well to simulations in both uniform and turbulent molecular clouds. In the following section we discuss how this compares to the simulations in this paper.

\subsection{Momentum from the Supernova Blastwave}
\label{aftersn:momentum}

\begin{figure*}
\centerline{\includegraphics[width=0.48\hsize]{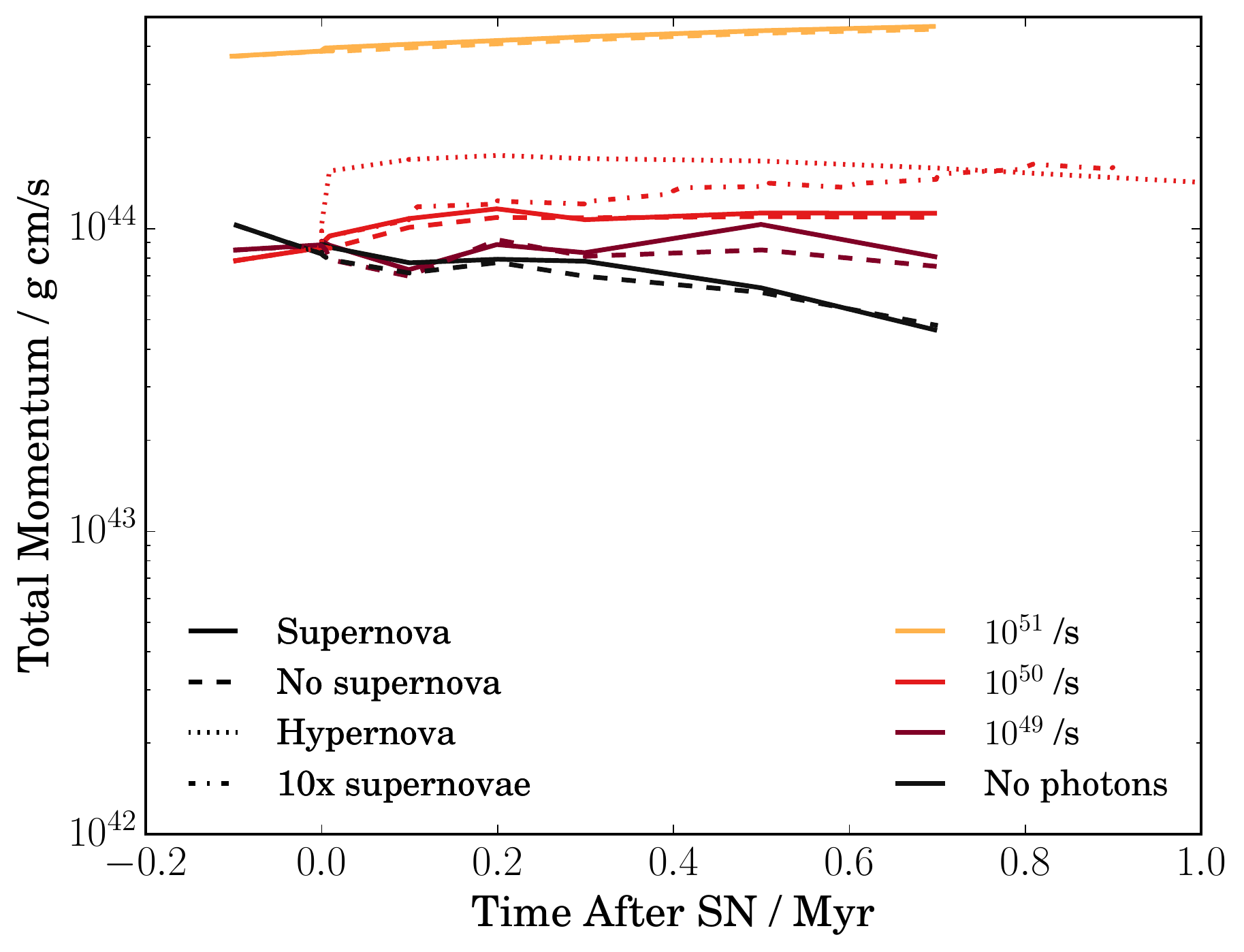}
\includegraphics[width=0.48\hsize]{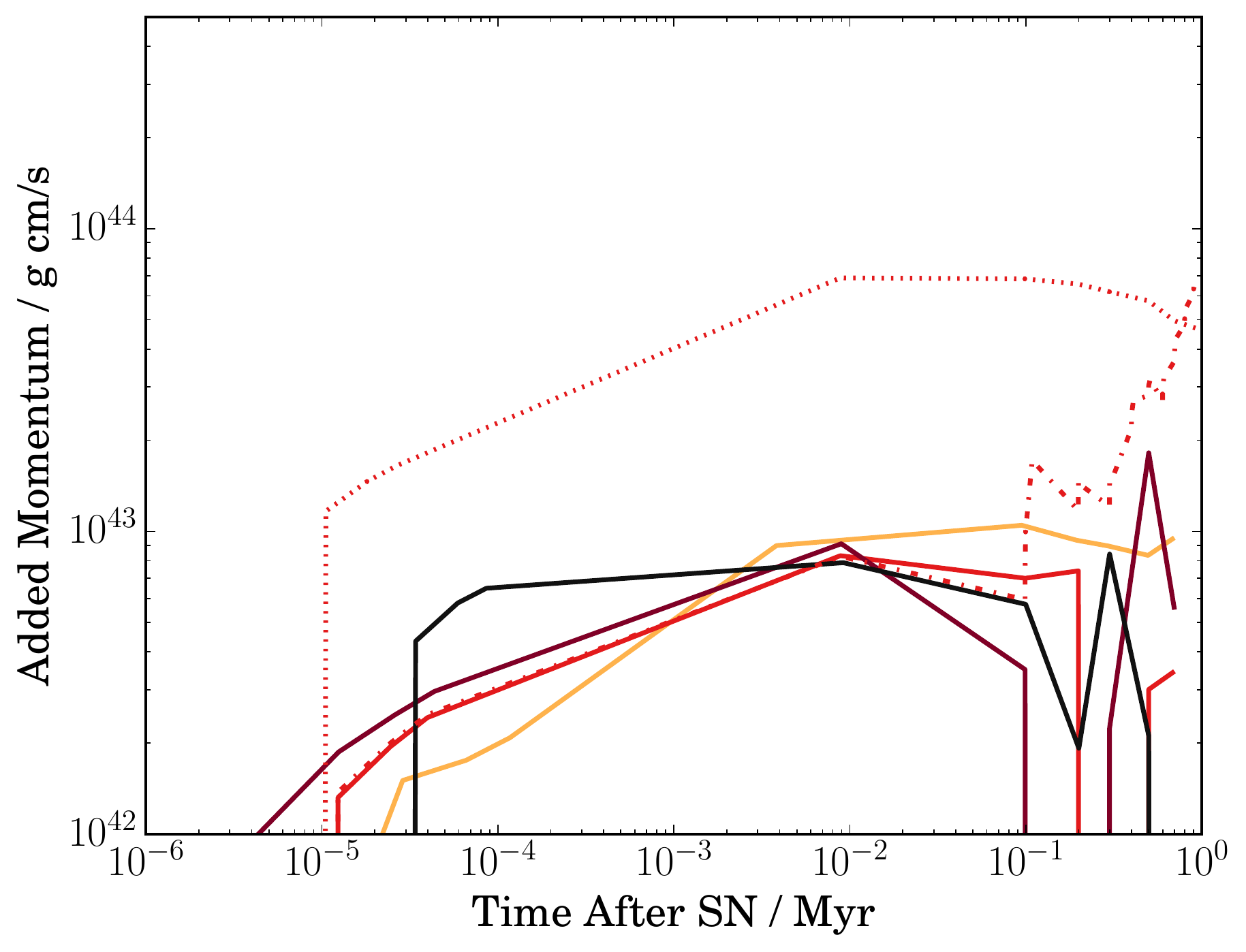}}
\caption{Momentum in radial flows in simulations with and without supernova feedback. On the left is a plot showing the total radial momentum with and without a supernova included, as well as the hypernova (``HN'') and multiple supernova (``MSN'') phase. The right-hand plot shows estimates of the momentum added by the supernova blaswaves in each simulation. These are made by taking the momentum at a given time in a simulation containing one or more supernovae or hypernova, and subtracting the momentum at the same time in an analogous run without a supernova. The time in the right-hand plot uses a log scaling to emphasise the early evolution of the blastwave. The $10^{51}$ photons/s results are taken from a simulation with twice the box length of the other simulations in order to prevent losing mass and momentum from the supernova shocks.}
 \label{aftersn:momentumfigure}
\end{figure*}

The supernova adds around $10^{43}$ g cm/s of radial momentum to the system, roughly an order of magnitude lower than the total momentum in flows in the cloud (see Figure \ref{aftersn:momentumfigure}). This estimate is made by subtracting the momentum in each simulation with a supernova from the momentum in an identical simulation without a supernova. Nonlinearities will add some error to the precise value found. 

Most of the momentum in the system is found in gas above 100 \atcc, with the exception of the $10^{51}$ photons/s simulation, which has succeeded in destroying the cloud. When ionising radiation is added, the supernova is able to add some momentum in the phase between 10 and 100 \atcc. Since the supernova leaves the adiabatic phase long before it escapes the cloud, its role is largely to accelerate clumps of gas away from the cloud rather than driving hot, diffuse winds out of the cloud.

We compare our results to estimates given in previous papers assuming the density at the position of the supernova, though there are strong density gradients around the position of the cloud (see Figure \ref{aftersn:densprofiles}). Estimates by \cite{Iffrig2015,Kim2015} give somewhere between 1.1 and $1.7 \times 10^{43}$ g cm/s of momentum for a cloud of density $10^3$ to $10^4$ \atcc, the densities found in the cloud around the supernova in all runs except when a $10^{51}$ photons/s source is included. Here, the density at the position of the source is closer to 100 \atcc, although even here we find energy losses due to interactions with dense cloud material as the supernova remnant expands. By contrast, \cite{Iffrig2015}, who use a similar setup with a $10^4$ \Msolar cloud, find that their results agree well with simulations performed in a uniform box with negligible external pressure forces.

Our momentum is lower than the values found by \cite{Walch2015}, who use a similar cloud mass, although our cloud is more fragmented and turbulent. We posit that this is because we pre-evolve our cloud with an initial turbulent field, whereas \cite{Walch2015} impose a fractal density field with zero initial velocity field. Both our simulations and \cite{Walch2015} include gravity. Tests performed in a uniform medium reproduce other works more closely, although a small difference is found due to the use of non-equilibrium hydrogen and helium cooling (Rosdahl et al, in prep.).

We suggest that the lower momentum found in our simulations is due to the density and velocity structure of the cloud. Our cloud fragments into dense, turbulent clumps as it evolves. These clumps have a large amount of ram pressure that opposes shocks interacting with them. For example, a parcel of fluid with density $10^4$ \atcc moving at 5 km/s provides $3 \times 10^{-9}$ ergs/cm$^3$ in ram pressure. The thermal pressure in the hot bubble of the supernova remnant is likely to be lower than this: $2\times10^{-9}$ ergs/cm$^{3}$ for a monatomic gas at $10^7$ K and 1 \atcc, which is a very high estimate for the density inside the hot bubble. In addition to this, \cite{McKee1977a} argue that the evaporation of dense clumps inside the hot bubble further reduces the energy of the supernova remnant.

\cite{Cioffi1988} argue that at the end of the supernova remnant's life, the shock will merge with the surrounding material. Equation 4.5 in that paper gives the time that this happens as:
\begin{equation}
 t_{merge} = 7.6 n_0^{-18/49} \epsilon_0^{31/98} \times 10^5 ~ \mathrm{years}
 \label{tmerge}
\end{equation}
 assuming the characteristic velocity of the turbulence is around $10~$km/s, which is to within an order of magnitude the value in our cloud. For $n_0$ = 100 and 10000 (in units of \atcc as in Equation \ref{aftersn:cox1972}), $t_{merge}$ = 0.14 and 0.026 Myr respectively. This is consistent with the transition in our simulations where the added radial momentum from the supernova remnant begins to fluctuate as radial flows are transferred to turbulent motions in dense clumps. The only simulation where this does not happen is in the $10^{51}$ photons/s simulation, where the cloud prior to the blast is highly porous (see Figure \ref{aftersn:pictures}).

 We discount the possibility that the shell around the supernova remnant decelerates due to gravity from material still embedded inside the cloud after the blastwave has expanded.  In Appendix \ref{appendix:shellgravity}, we calculate that this should only become important for supernova remnants moving through densities above $10^3$ \atcc. Since the majority of the cloud is at a lower density, we expect this to only be a secondary effect for this system.

\section{Beyond $10^{51}$ ergs}
\label{moresne}

The previously discussed results were for the case where a single supernova with energy $10^{51}$ ergs was injected. In this Section we discuss two simple extensions to this model that include both a rare but powerful hypernova explosion, and a simple model of multiple supernovae exploding in the same cluster.

\subsection{Hypernovae}
\label{hypernovae}

According to \cite{Heger2003} above around 25 \Msolar stars become black holes rather than neutron stars, leading to weak supernovae. However, \cite{Nomoto2005} find that a small fraction of very massive stars with high rotation rates will explode as hypernovae, which inject 10 times or more the energy of typical supernovae. In this section we discuss a simulation with a $10^{50}$ /s source where we inject a hypernova of energy $10^{52}~$ ergs into the cloud. In Figure \ref{aftersn:momentumfigure} we find ten times the momentum from the hypernova as the standard supernova model, agreeing with Equation \ref{aftersn:pc}. In this case the blast adds sufficient momentum that the signal is not lost in the turbulent motions of the cloud.

The total momentum in flows balancing the gravitational forces in the cloud is approximately the cloud mass multiplied by the virial velocity at the cloud edge. For a cloud of $10^5$ \Msolar with a total radius of 20 pc (see Figure \ref{aftersn:densprofiles}) this is roughly $10^{44}$ g cm / s, approximately the amount of momentum despoited by the hypernova. A hypernova thus deposits sufficient momentum to counteract the gravitational forces binding the cloud. In practice, due to the clumpy nature of the medium and imperfect coupling of the hypernova blastwave to the clumps, we find that some cloud material is able to remain embedded in the cloud.

\subsection{Multiple Supernovae}
\label{multiplesne}

In cloud of $10^5$ \Msolar there will typically be multiple massive stars in the embedded cluster. There will hence be multiple supernovae that explode in the same cloud. In this Section we analyse a simulation in which we inject nine further $10^{51}$ erg supernovae every 0.1 Myr after the first supernova. This is similar to the experiment performed by \cite{Kim2015} in a turbulent environment. Each time a supernova is injected it occurs in an environment swept out by the previous supernovae, gradually inflating the supernova remnant in short bursts.

In Figure \ref{aftersn:momentumfigure} we find that the momentum injected by these supernovae gives roughly the same end result as the hypernova. Again, this is consistent with Equation \ref{aftersn:pc}, in which the final momentum is roughly proportional to the energy of the blast. We do not find the large increase in momentum with subsequent supernovae as \cite{Gentry2016} do in their 1D study, despite reaching a comparable spatial resolution (0.084 pc compared to 0.06 pc). Part of this is due to the fact that dense material remains embedded in the cloud even after the initial supernova. However, the claim in \cite{Gentry2016} that Eulerian codes suffer from overmixing near the shell boundary should be carefully studied, particularly in combination with 3D turbulence and full non-equilibrium radiative cooling.

For a cluster with mass $10^4$ \Msolar (10\% of the total mass of the cloud), using an \IMF slope of -2.35 in the high mass end we estimate 160 stars above 8 \Msolar. In the range 4 to 20 Myr, this gives a mean time delay of 0.1 Myr, although due to the shape of the \IMF more stars will explode at later times. We thus expect this model to be an underestimate for the total injection of energy by supernovae into the cloud by a factor of 16. However, if stars form over around $t_{ff}$ in the cloud (2.53 Myr), these supernovae will occur too late to prevent star formation. Rather, they will be more effective at dispersing any remaining cloud material.

\section{Discussion}
\label{discussion}

In this work we have focussed on the combined role of ionising radiation and supernovae in destroying molecular clouds, and the interaction between the two processes. However, there are a number of important aspects still to be explored in both reproducing and explaining star formation and feedback in molecular clouds.

A number of physical processes have not been included in this work. Radiation pressure is omitted, though there is some debate in the literature regarding its effectiveness. \cite{Agertz2013,Hopkins2014} assume strong coupling between infrared photons and dust grains, while \cite{Krumholz2012c,Sales2014,Rosdahl2015a,Haworth2015} argue that radiation pressure has a minimal effect in the regimes studied.

Stellar winds too are omitted. \cite{Dale2014} argue that winds are not likely to be as effective at driving outflows as ionising radiation, although these authors neglect late stage stellar evolution where winds become stronger. This source of late-stage feedback is particularly important because the amount of UV photons drops as the star expands and cools. Corrections by \cite{Kudritzki2000} for the wind velocity boost wind energies by up to a factor of 10 over values assuming a wind travelling at the escape velocity of the star. By combining recent models of stellar evolution by \cite{Ekstrom2012} with the boost from \cite{Kudritzki2000}, cumulative energies from stellar winds can exceed that from a supernova, though this energy will be spread out across the lifetime of the star rather than as a single burst of energy as in a supernova.

\cite{Mason2009,DeMink2014} find that the majority of observed massive stars are in binaries. As such many of them are rotating and winds may become more efficient in this regime. This means that as these stars lose their envelopes they stay hot in the final stage of their evolution, producing more UV photons \citep{Kohler2014}.

The model for supernovae becomes more complex for very massive stars. \cite{Heger2003,Nomoto2005} discuss the fate of stars above 25 \Msolar, which can become either weak supernovae or very energetic hypernovae. \cite{Podsiadlowski2004} argue that the latter events are rare, though since they can deposit significantly more energy into the \ISM than $10^{51}$ ergs, they may be significant events, as we find in Section \ref{hypernovae}.

Finally, processes such as protostellar jets \citep[see review by ][]{Frank2014} and x-ray emission from stellar-mass black holes \citep{Mapelli2014} can also increase the energy budget for stellar feedback in the cloud. For a review of stellar feedback processes in star-forming regions, see \cite{Dale2015a}.

The time over which the star emits energy in various forms depends strongly on the stellar evolution models used. In this paper we assume a lifetime of 3 Myr followed by a supernova for the most massive stars. We omit a period of around 1 Myr during which stellar winds, which we do not include in this paper, become important. Binary evolution, which affects a majority of observed very massive stars, also affects the lifetime of stellar winds and ionising radiation from the stars. There are thus many open questions regarding how much energy is available from feedback in various forms, and over which timescale.

Star formation in the cloud and the response of the cloud to radiation both occur on a scale of a free-fall time. There is hence a competition between the two processes that will, in part, set the star formation efficiency of the cloud. Since in, e.g., \cite{Matzner2002}, the expansion of the ionisation front is proportional to $S_*^{1/7}$, the cloud is not highly sensitive to the precise number of photons produced by the cluster. Additionally, stars form in very dense regions that are underresolved by our simulations, rather than at the centre of the cloud as in this work. We will begin to address these question in future work.

In our simulations we find that a $10^{51}$ ergs supernova will add $10^{43}$ g cm/s of momentum to a $10^5$ \Msolar cloud, mostly in the dense phase. Adding more energy, either in a single hypernova or as multiple supernovae, adds proportionally more momentum. Eventually enough momentum is added to unbind the cloud. 

The role of supernovae in setting the star formation efficiency of the cloud is unclear. Since the first supernovae occur a few Myr after the first star is formed, they cannot immediately regulate star formation in the cloud. In addition, one supernova will not be enough to unbind a massive cloud. By contrast, ionising radiation is capable of disrupting star-forming clouds, particularly in the case studied here where a cluster producing $10^{51}$ photons/s unbinds the entire cloud prior to the first supernovae. In this scenario, the supernovae will be injected directly into the diffuse medium.

There are two advantages to invoking supernova feedback in a cloud environment as opposed to ionising radiation. The first is that they couple directly to the gas, and as such are more efficient at transferring their energy to the gas than ionising photons. \cite{Walch2012} estimate an efficiency of energy from ionising photons to kinetic energy in the gas as approximatedly 0.1\%, compared to the few percent found by, e.g., \cite{Chevalier1974} for supernovae. The second is that they are capable of driving outflows at high velocities, whereas photoionised gas can only expand at around 10 km/s.

Different environments will have different behaviours. For example, high redshift HII regions will have higher temperatures since metal cooling is absent, but they will be embedded in denser, higher pressure environments. Very massive clouds will also be difficult to disperse. For example, \cite{Krause2016} find that only a large number of hypernovae are able to prevent the majority of a massive extragalactic cloud from turning into stars. More work must be done to extend the findings of this paper in local environments to more universal conditions.

\section{Conclusions}
\label{conclusions}

We perform a series of radiative magnetohydroynamic simulations of a $10^5$ \Msolar turbulent molecular cloud with embedded sources of ionising radiation and a supernova. We compare the results of these simulations to analytic models given in the previous paper and find good agreement provided a good fit is found for the density field of the cloud. We study the time evolution of these analytic models, finding that the limit at which the ionisation front stalls (i.e. stops expanding) is reached over approximately one free-fall time in the cloud. This introduces a competition between the two processes of star formation and stellar feedback.

The environment that the supernova blastwave expands into depends strongly on the emission rate of ionising photons from the cluster beforehand. An emission rate of $10^{51}$ photons per second, roughly equivalent to a cluster of $10^4$ \Msolar or 10\% of the total cloud mass, is capable of disrupting the cloud, though dense clumps remain. If the number of stars formed is much lower, the ionising photons will not be able to destroy the cloud and the supernova will transfer its momentum to the dense cloud gas rather than the diffuse interstellar medium. This suggests that for this cloud a star formation efficiency of approximately 10\% is expected if the main source of feedback is from ionising photons. The position of the source of ionising photons and the supernova is highly important due to the effect of gas density on photon recombination and radiative cooling in general.

We inject supernovae as a single thermal pulse of $10^{51}$ ergs. We also perform simulations with ten supernovae of the same energy 0.1 Myr apart or one hypernova of $10^{52}$ ergs. The resulting total momentum from our supernovae is roughly $10^{43}$ g cm/s per $10^{51}$ ergs of injected energy. This is at the low end of the values given in other works, but it is not inconsistent provided the early evolution of the blastwave occurs in gas at around $10^4$ \atcc or higher. We argue that flows of dense, turbulent gas inside the cloud are capable of reducing the momentum added to the \ISM by supernovae as long as dense clumps remain embedded within the cloud at the time the supernova occurs. Most of the momentum from a single supernova is deposited into the dense gas rather than as fast, hot, diffuse flows, except in cases where the ionising photons have swept away most of the cloud.

 We speculate that supernovae will occur too late to prevent the bulk of star formation in the cloud, but sufficient supernovae will be capable of expelling the remaining gas and allowing future supernovae to drive shocks into the interstellar medium.

\section{Acknowlegements}
\label{acknowledgements}

We would like to thank Olivier Iffrig, Romain Teyssier, Andreas Bleuler, Alex Richings, Suzanne Madden, Yueh-Ning Lee, for useful discussions during the preparation of this paper. The simulations presented here were run on the machine irfucoast at CEA Saclay. This work has been funded by the the European Research Council under the European Community's Seventh Framework Programme (FP7/2007-2013). SG and PH are funded by Grant Agreement no. 306483 of this programme. JR is funded by Grant Agreement 278594-GasAroundGalaxies of the same programme and the Marie Curie Training Network CosmoComp (PITN-GA-2009-238356).

 \bibliographystyle{mnras}
 \bibliography{hiisnpaper}

\appendix

\section{Emission from Clusters}
\label{appendix:clusteremission}

\begin{figure}
\centerline{\includegraphics[width=0.98\hsize]{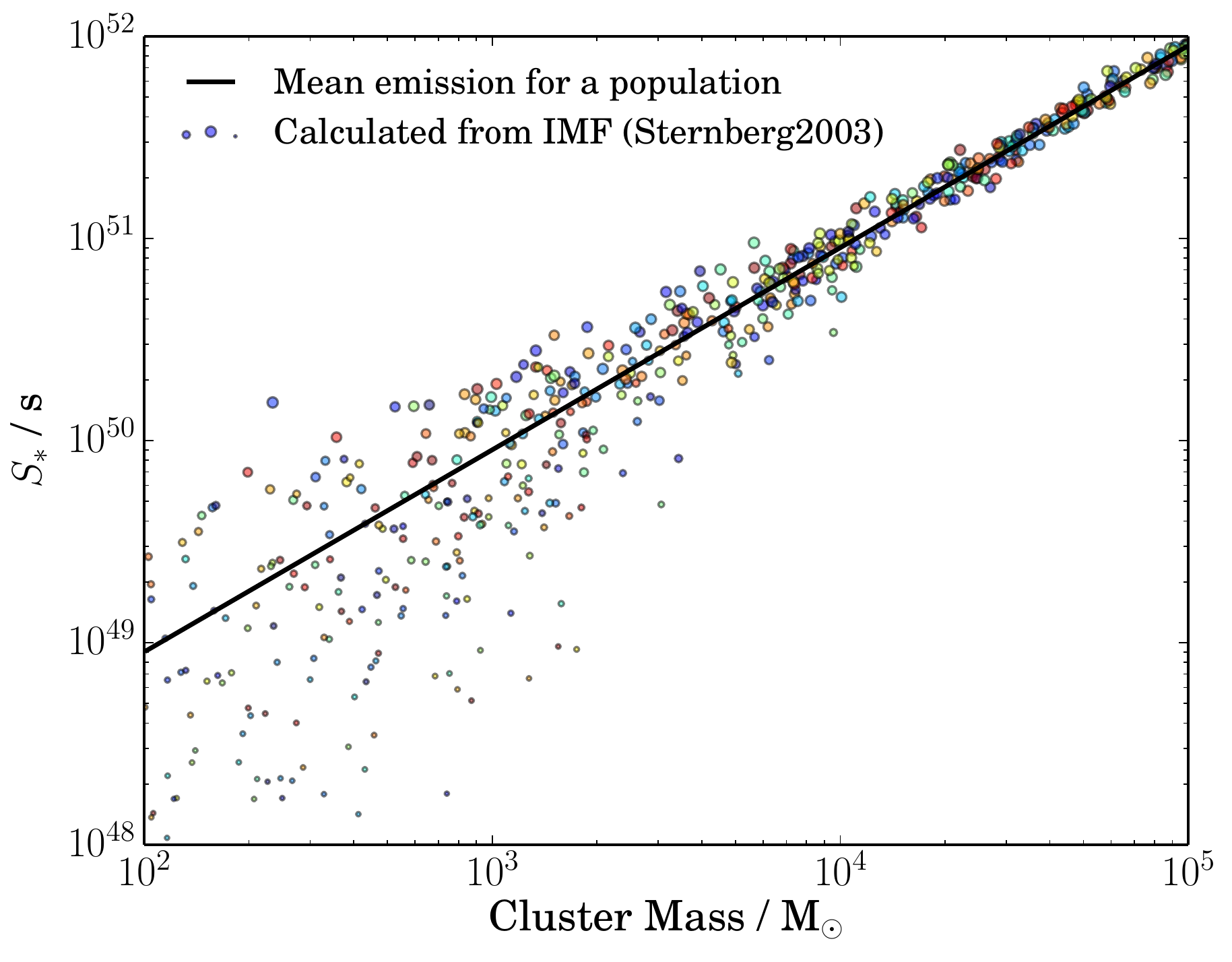}}
\caption{Hydrogen-ionising photon emission rate for clusters of various masses. The scatter plot shows values calculated by sampling an IMF \citep[e.g ][]{Chabrier2003} and assigning photon emission rates to each star using the values given by \protect\cite{Sternberg2003}. The size of the points is proportional to the most massive star in the cluster. The solid line is given by $S_* = k M_*$, where $M_*$ is the cluster mass, and $k$ is found by calculating $ \sum S_{*} / \sum M_{*}$ for each cluster.}
 \label{clusteremission:clusterQH}
\end{figure}

In this section we estimate the approximate hydrogen-ionising photon emission rate from star clusters of various masses. We sample stellar masses for a set of clusters of various total masses using a random Monte Carlo sampling. We do not a priori force the maximum stellar mass to be below a maximum value given by, e.g., \cite{Weidner2009}. Ionising hydrogen fluxes for each stellar mass are found by interpolating the results of \cite{Sternberg2003}, though the results do not differ strongly from values found using earlier work by \cite{Vacca1996}. We plot these results in Figure \ref{clusteremission:clusterQH}, along with a fit assuming that the cluster is a perfectly sampled stellar population. Below a few thousand \Msolar the IMF is incompletely sampled, and as such statistical noise begins to cause a large spread in the results. However, the linear fit is still reasonable given the large spread of photon emission rates.

\section{Expansion Solution for the Ionisation Front}
\label{appendix:expansion}

The spherically-averaged expansion rate of the ionisation front with radius $r_i$ at time $t$ in a turbulent cloud, as derived in \cite{Geen2015b} and based on \cite{Raga2012}, is given by
\begin{equation}
\frac{1}{c_i}\frac{\mathrm{d}r_i}{\mathrm{d}t} = F(r,t)-\frac{c_{ext}^2}{c_i^2}\frac{1}{F(r,t)}+\frac{v_{ext}(r,t)}{c_i} ,
\label{expansion:raga_like}
\end{equation} where
\begin{equation}
F(r,t) \equiv \sqrt{\frac{n_i}{n_{ext}}} = \left(\frac{r_{s}}{r_i}\right)^{3/4}\left(\frac{n_c}{n_{ext}(r,t)}\right)^{1/2} .
 \label{expansion:raga_term}
 \end{equation} $c_{ext}$ is a term including the sound speed and turbulent motions in the gas just outside the shock radius and $v_{ext}$ is the velocity of the gas just outside the shock radius normal to the shock surface (assumed in 1D to be radial from the source position). $c_i$ is the sound speed in the ionised gas and $n_i$ is the density in the ionised gas. $r_{s}$ is the initial \Stromgren radius, i.e. the radius at which the ionisation front reaches equilibrium assuming a hydrostatic approximation. $n_c$ and $n_{ext}$ are as defined in Equation  \ref{beforesn:density_profile}.

\section{Timescale for Ionisation Front Expansion}
\label{appendix:timevolution}

In this section we provide a simplistic calculation for the typical timescale for the expansion of the ionisation front in Equation \ref{expansion:raga_like} as it approaches $r_{stall}$, where $\dot{r} = 0$. We make the simplifying assumption that $v_{ext} \gg c_{ext}$ (a solution assuming the reverse would be equally valid). We solve this equation for the flat core in the cloud, where the $n_{ext} = n_0$ is constant.

\begin{figure}
\centerline{\includegraphics[width=0.98\hsize]{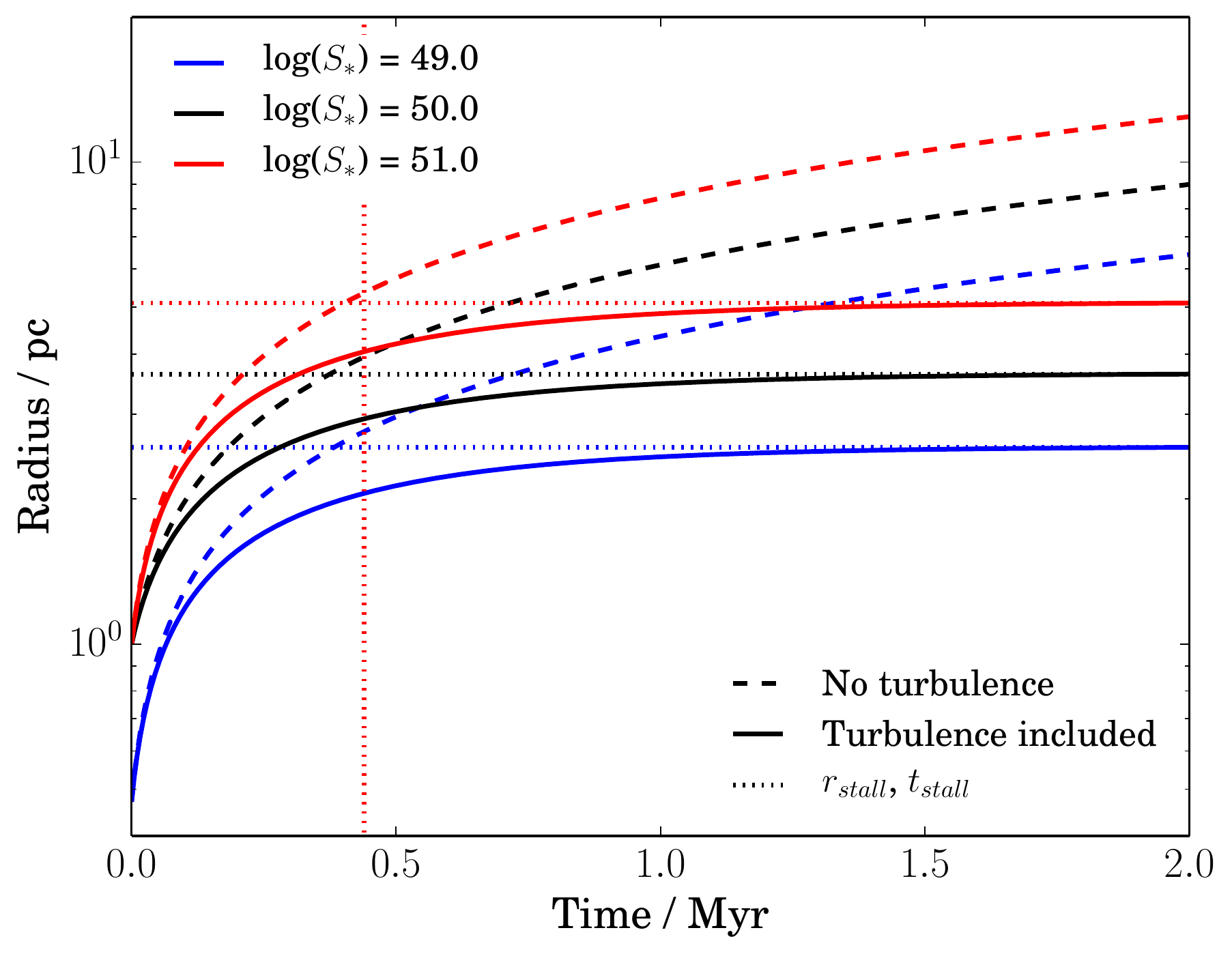}}
\caption{Numerical solutions for the radial evolution of the ionisation front in Equation \ref{expansion:raga_like} assuming cloud properties as given in Section \ref{beforesn}. Overplotted are values for $r_{stall}$ and $t_{stall}$. Note that since Equation \ref{te:free} makes the assumption that $r_s$ is reached over a negligible time, the match between the numerical solution and the intersection of $r_{stall}$ and $t_{stall}$ is not exact.}
 \label{timeevolution:rstall}
\end{figure}

A full solution of these equations requires hypergeometric functions, which are difficult to interpret. Instead, we adopt the following simplistic estimate for the timescale over which the ionisation front stalls. This is taken to be the time that the solution to this equation assuming that $v_{ext} = 0$ \citep{SpitzerLyman1978,Matzner2002} reaches the same radius as $r_{stall}$. The reason we do this is that it provides a reasonable first order estimate for the time at which $v_{ext}$ becomes a limiting factor in the expansion of the ionisation front. We compare this time to numerical solutions to Equation \ref{expansion:raga_like} in Figure \ref{timeevolution:rstall}. 

If $v_{ext} = 0$, and assuming the ionisation front rapidly reaches $r_s$ (i.e. the recomination time is negligible) we can write
\begin{equation}
 r_i(t) = r_s \left( \frac{7}{4} \frac{c_i t}{r_s} \right)^{\frac{4}{7}}.
 \label{te:free}
\end{equation}
Alternatively, if $v_{ext}$ is non-negligible, $r_i$ tends towards a limit where $\dot{r} = 0$
\begin{equation}
 r_{stall} = r_s \left( \frac{c_i}{v_{ext}} \right )^{\frac{4}{3}}.
 \label{te:stall}
\end{equation}
Setting $r_i(t_{stall}) = r_{stall}$, and combining Equations \ref{te:free} and \ref{te:stall}, we find
\begin{equation}
 \left( \frac{c_i}{v_{ext}} \right )^{\frac{4}{3}} = \left( \frac{7}{4} \frac{c_i t_{stall}}{r_s} \right)^{\frac{4}{7}}.
\end{equation} 
Invoking Equation \ref{te:stall} again to replace $r_s$ with $r_{stall}$, we find:
\begin{equation}
 t_{stall} = \frac{4}{7} \frac{r_{stall}}{v_{ext}}.
\label{te:time}
 \end{equation}
We now calculate the value of $r_{stall}/v_{ext}$. We assume, as in Section \ref{beforesn}, that $v_{ext}$ is the virial velocity at $r_{stall}$, i.e.
\begin{equation}
 v_{ext}^2 = \frac{6}{5}\frac{G M}{r_{stall}}
\end{equation}
where
\begin{equation}
 M = \frac{4}{3} \pi r_{stall}^3 \rho_0 
\end{equation}
where $\rho_0 = n_0 m_H / X$. The free-fall time in this cloud core is
\begin{equation}
 t_{ff} = \sqrt{\frac{3 \pi}{32 G \rho_0}}
\end{equation} and hence we can write
\begin{equation}
t_{ff} = \sqrt{\frac{6}{5}\frac{\pi^2}{8}} \frac{r_{stall}}{v_{ext}} = 1.2 \frac{r_{stall}}{v_{ext}}
\label{te:tff}
\end{equation}
Combining Equations \ref{te:tff} and \ref{te:time}, we can write
\begin{equation}
 t_{stall} \simeq 0.7 t_{ff}.
\end{equation}
Comparing this value as plotted on Figure \ref{timeevolution:rstall} (by eye), the ionisation front will reach a value close to $r_{stall}$ over $\simeq 2 t_{stall}$. Hence the timescale over which the ionisation front stalls, assuming it remains inside the core where $n_{ext} = n_0$, is roughly $1.4 t_{ff}$, which is on the order of $t_{ff}$. Note that this is a very crude estimate, as it simplifies greatly the full equations that govern the ionisation front. We discuss this result further in Section \ref{beforesn:comparison}.
 
\section{Collapsing Ionisation Fronts}
\label{appendix:collapsing}

In our simulation \simname{N49-NSN} the ionisation front collapses after expanding briefly. We introduce a simple spherically symmetric model invoking accretion onto the cloud core. \cite{Ntormousi2015}, based on \cite{Larson1969}, give the time-dependent density of an accreting cloud as
\begin{equation}
 n_{ext} = \frac{n_0}{\left(1 - \frac{t}{t_{ff}}\right )^2}
\end{equation}

Solving Equation \ref{expansion:raga_like} using this equation and assuming no external velocity field, we arrive at
\begin{equation}
r_i(t) = r_s \left\{ \frac{7}{4} \frac{c_i}{r_s} t \left(1-\frac{t}{2 t_{ff}}\right) \right\}^\frac{4}{7}
\label{beforesn:collapsesoln}
\end{equation}
Note that the density field has a singularity at $t_{ff}$. Beyond this point the solution for $r_i$ becomes unphysical, and we instead keep $r_i = 0$.

\begin{figure}
\centerline{\includegraphics[width=0.98\hsize]{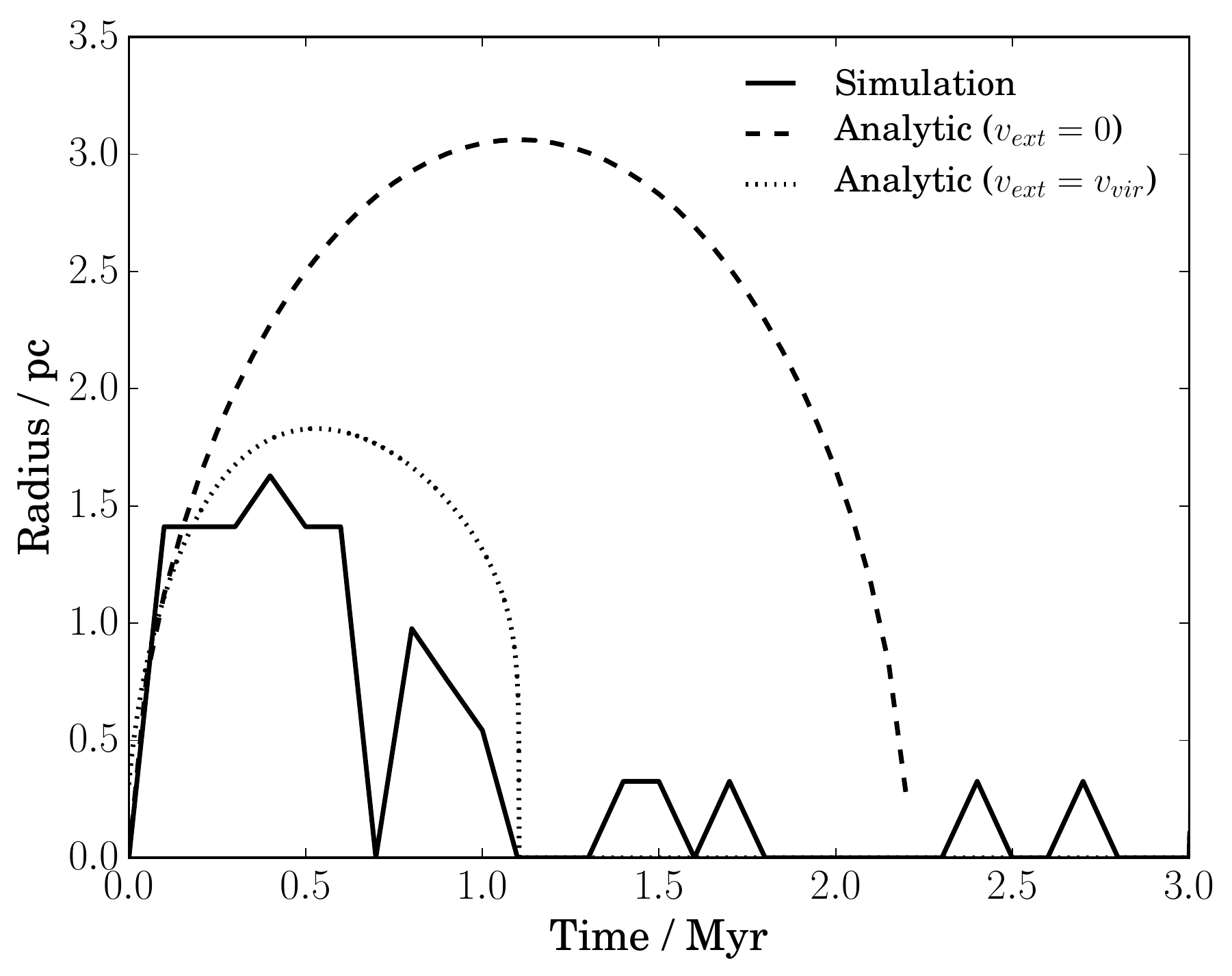}}
\caption{Comparison between the radial expansion and collapse of the ionsation front in the simulations and analytic models. The simulation is \protect\simname{N49-NSN}, i.e. using a $10^{49}$/s ionising photon source in the cloud discussed in this paper. Analytic solutions are found to Equation \protect\ref{expansion:raga_like} as described in Appendix \protect\ref{appendix:collapsing}.}
\label{beforesn:collapsingfig}
\end{figure}

We plot the solution to Equation \ref{beforesn:collapsesoln} in Figure \ref{beforesn:collapsingfig} for the $10^{49}$ /s source, whose ionisation front collapses in our cloud. In addition we solve Equation \ref{expansion:raga_like} numerically in the case where $v_{ext}$ is set to the virial velocity (as described in Section \ref{beforesn:comparison}). As in Section \ref{beforesn:comparison}, $t_{ff}$ is calculated for the cloud core rather than the cloud as a whole. We find that the numerical solution with $v_{ext} = v_{vir}$ agrees with the simulation results reasonably well, except for the flickering of the HII region due to the orbits of dense clumps passing through the source position. 

Hence in both the case where the front stalls and the case where it collapses, the time evolution of the ionisation front is governed by the free-fall time in the cloud core. 

\section{Shell Expansion Under Gravity}
\label{appendix:shellgravity}

In a spherically-symmetric solution, the supernova must entrain all the material in its path as it expands. However, in the 3D case, dense clumps of gas or (not included in these simulations) stars will remain embedded inside the supernova remnant. This provides an additional gravitational force on the shell. Similar models are derived for HII regions in \cite{Garcia-Segura1996,Keto2002,Didelon2015}.

In this model we assume a shell moving outwards spherically, entraining all mass enclosed within it except for a fixed central mass. The mass of the shell is assumed to be the total mass $m(r)$ displaced by the shell at $r$. We assume a power law density field with index $w$ and characteristic density $\rho_0$ and radius $r_0$, defined as $\rho(r) = \rho_0 {r/r_0}^{-w}$. Integrating, we find $m(r) = \frac{4}{3} \pi \rho_0 r_0^w r^{3-w}$. The reaction force from mass accretion onto the shell in a power law density field can be written as:
\begin{equation}
 m\ddot{r} = -\dot{m}\dot{r} = -\frac{\mathrm{d} m}{\mathrm{d} r}\dot{r}^2.
\end{equation}
Dividing by $m$, we find
\begin{equation}
 \ddot{r} = -\frac{(3-w) \dot{r}^2}{r}.
\end{equation} 
Note that this equation assumes the mass outside the supernova remnant is static. In our simulations this assumption is not too unreasonable, since the cloud is roughly virialised and the dissipation timescale for the turbulence in the cloud that drives dynamic evolution in the cloud is longer than the time over which the supernova remnant evolves.

Including gravity, the equation becomes 
\begin{equation}
 \ddot{r} = -\frac{G M_c}{r^2} = -\frac{(3-w) \dot{r}^2}{r}.
 \label{aftersn:gravdeceleration}
\end{equation} where the central mass is $M_c$. Note that neither equation depends on the ambient density of the medium, although the density will set the initial velocity and radius of the shell as it enters the momentum-driven phase as given in Equation \ref{aftersn:pc}.

Note that if $r$ shrinks, $M_c$ will also drop, whereas in reality a contracting shell would retain its mass. Thus the solution to Equation \ref{aftersn:gravdeceleration} after $r$ begins shrinking should be used with some caution. Rather, the value of this expression is determining at what point the shell stalls under gravity in the presence of a central gravitating mass.

\begin{figure}
\centerline{\includegraphics[width=0.98\hsize]{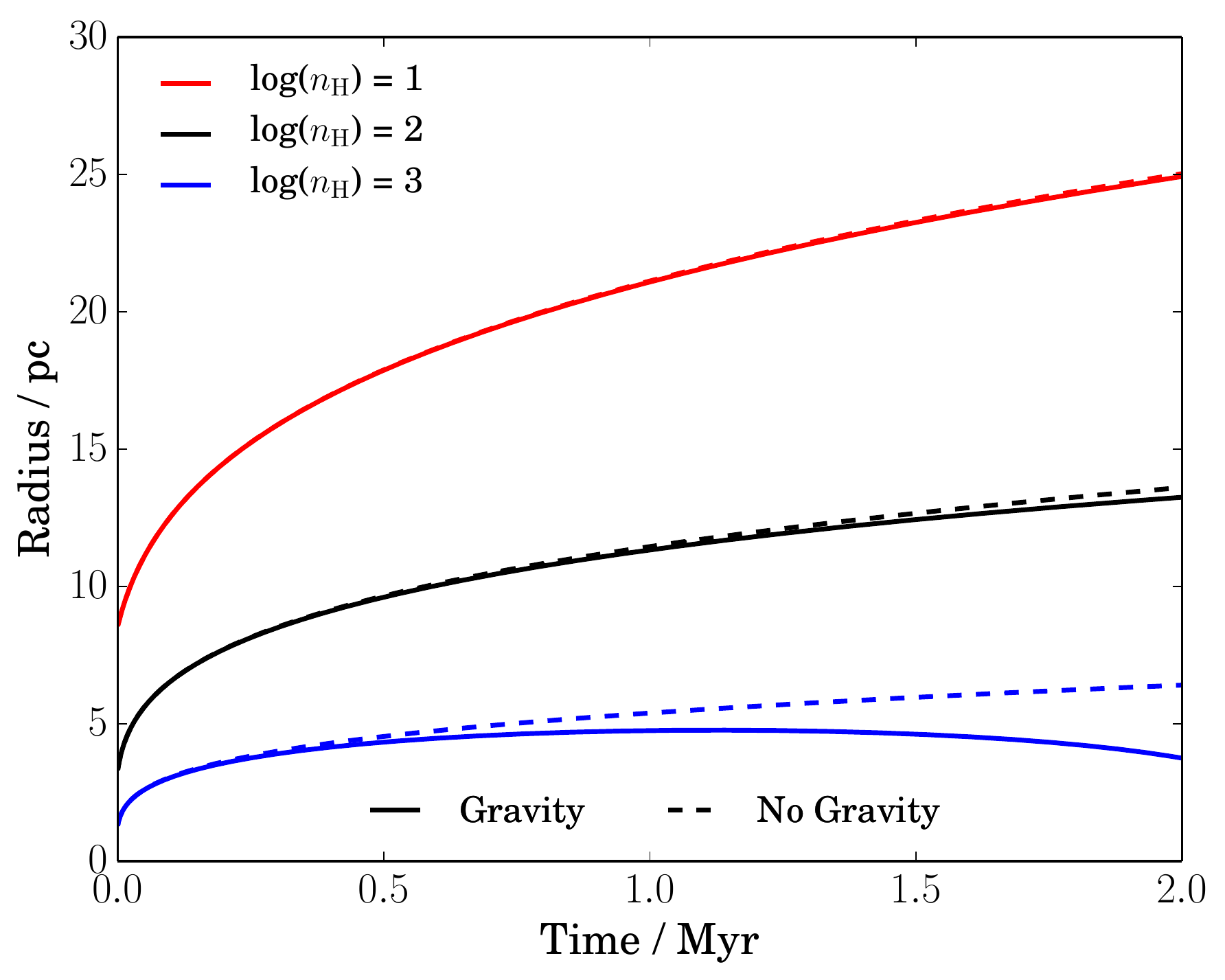}}
\caption{Solutions to Equation \protect\ref{aftersn:gravdeceleration} in Section \protect\ref{appendix:shellgravity} with and without gravity. We assume an initial ballistic momentum $10^{43}$ g cm/s, radius $r_c$ in Equation \protect\ref{aftersn:cox1972} and $M_c=10^4$ \Msolar.}
\label{aftersn:gravsolutions}
\end{figure} 

In Figure \ref{aftersn:gravsolutions} we provide numerical solutions for this equation with and without the gravity for three values of $\rho_0$ with $w=0$ and an initial shell velocity and radius calculated by assuming a momentum of $10^{43}$ g cm/s and radius $r_\Lambda$ (Equation \ref{aftersn:cox1972}). The mass of the embedded cluster $M_c$ is assumed to be a point mass of $10^4$ \Msolar, or 10\% of the total mass of the cloud. In our simulations, gravity should only become important above $10^3$ \atcc. This is a highly simplistic view of the properties of the central mass, so these results should be considered largely illustrative.

\let\urshowkeys=\showkeys \def\showkeys{\needspace{5ex}\urshowkeys}

\end{document}